\documentclass[preprint,,showpacs,preprintnumbers,amsmath,amssymb]{revtex4-1}

\usepackage{graphicx}
\usepackage{epstopdf}
\usepackage{dcolumn}
\usepackage{bm}
\usepackage{amsmath}
\usepackage{graphicx} 
\usepackage{dcolumn}
\usepackage{bm}
\usepackage{amsmath}
\usepackage{epsfig}
\usepackage{float}
\usepackage{epstopdf}
\begin{document}

\title{Ultrafast Control of Excitonic Rashba Fine Structure by Phonon Coherences in a Metal Halide Perovskite CH$_3$NH$_3$PbI$_3$}

\author
{Z. Liu$^{1\ast}$, C. Vaswani$^{1\ast}$, X. Yang$^{1}$, X. Zhao$^{1}$, Y. Yao$^{1}$, Z. Song$^{2}$, D. Cheng$^{1}$, Y. Shi$^{3}$, L. Luo$^{1}$, D. -H. Mudiyanselage$^{1}$, C. Huang$^{1}$, J.-M. Park $^{1}$, R.H.J. Kim $^{1}$, J. Zhao$^{3}$, Y. Yan$^{2}$, K.-M. Ho$^{1}$,
	J. Wang$^{1\dag}$}
\affiliation{$^1$Department of Physics and Astronomy and Ames Laboratory-U.S. DOE, Iowa State University, Ames, Iowa 50011, USA. 
	\\$^2$Department of Physics and Astronomy and Wright Center for Photovoltaics Innovation and Commercialization, The University of Toledo, Toledo, OH 43606, USA.
\\$^3$ICQD/Hefei National Laboratory for Physical Sciences at the Microscale, and Key Laboratory of Strongly-Coupled Quantum Matter Physics, Chinese Academy of Sciences, University of Science and Technology of China, Hefei, Anhui 230026, China}

\date{\today}

\begin{abstract}
We discover hidden Rashba fine structure in CH$_3$NH$_3$PbI$_3$ and demonstrate its quantum control by vibrational coherence through symmetry-selective vibronic (electron-phonon) coupling. Above a critical threshold of a single-cycle terahertz pump field, a Raman phonon mode distinctly modulates the middle excitonic states with {\em persistent} coherence for more than ten times longer than the ones on two sides that predominately couple to infrared phonons.  These vibronic quantum beats, together with first-principles modeling of phonon periodically modulated Rashba parameters, identify a {\em three-fold} excitonic fine structure splitting, i.e., optically-forbidden, degenerate dark states in between two bright ones. 
Harnessing of vibronic quantum coherence and symmetry inspires light-perovskite quantum control and sub-THz-cycle ``Rashba engineering" of spin-split bands for ultimate multi-function device.    
\end{abstract}
\maketitle
Fundamental understanding and light quantum control of metal halide perovskites are key to discovering new photovoltaic materials and multi-function spin-charge-photon devices based on methylammonium lead iodide, MAPbI$_3$, illustrated in Fig. 1(a). 
Rashba-type effects, which control the direct or indirect nature of spin-split bands and fine structure splitting (FSS) \cite{Stranks,Becker},
have been proposed to determine the outstanding, yet mysterious properties in perovskites \cite{PRB}. These include, e.g., long photocarrier lifetimes \cite{Zheng}, spin/charge diffusion lengths \cite{Giovanni, spin} and nonlinear optics \cite{K}. 
The challenge to unambiguously detect the FSS in them lies in the vibronic fluctuations on short time and length scales associated
with dynamical entropy and local symmetry breaking that seriously broaden band edge absorption. 
As illustrated in the Fig. 1(b) (inset) \cite{Isarov}, the Rashba effects lift spin and momentum degeneracies in MAPbI$_3$ which leads to the excitonic FSS, where the four fold degeneracy of the 1{\em s} exciton is split by the Rashba terms $\alpha _{e(h)} {{\vec  \sigma_{e(h)}}}\cdot \hat{n}\times i\vec{\bigtriangledown}_{r_{e(h)}}$. 
Here $\alpha _{e}$($\alpha _{h}$) and $\sigma_{e}$($\sigma_{h}$) denote the Rashba parameter and Pauli matrices for the $\mathbf J$=1/2 ($\mathbf S$=1/2) conduction (valance) band, respectively.  
Energy levels of the lowest-lying excitonic fine structure can be calculated based on these parameters \cite{Isarov} (supplementary) as two bright exciton states (green lines), $\left | \Phi _{1,4}\right \rangle$, lying above and below two degenerate dark states (blue lines), $\left | \Phi _{2,3}\right \rangle$, in MAPbI$_3$ (inset, Fig. 1(b)).  
Consequently the bight and dark exciton splitting is only few nm,
$\sim$10 times smaller than the inhomogeneously broadened PL linewidth ($>$30 nm) \cite{Liu, Beecher}. 
The presence of the FSS and dark states, the hallmark for the Rashaba effects in perovskites, remain elusive until this work despite intense studies \cite{Innocenzo,Milot,Liu, Tom, Isarov,Fu, wilhelm,JPC}. 

\begin{figure}
	\begin{center}
		\includegraphics[width=150mm]{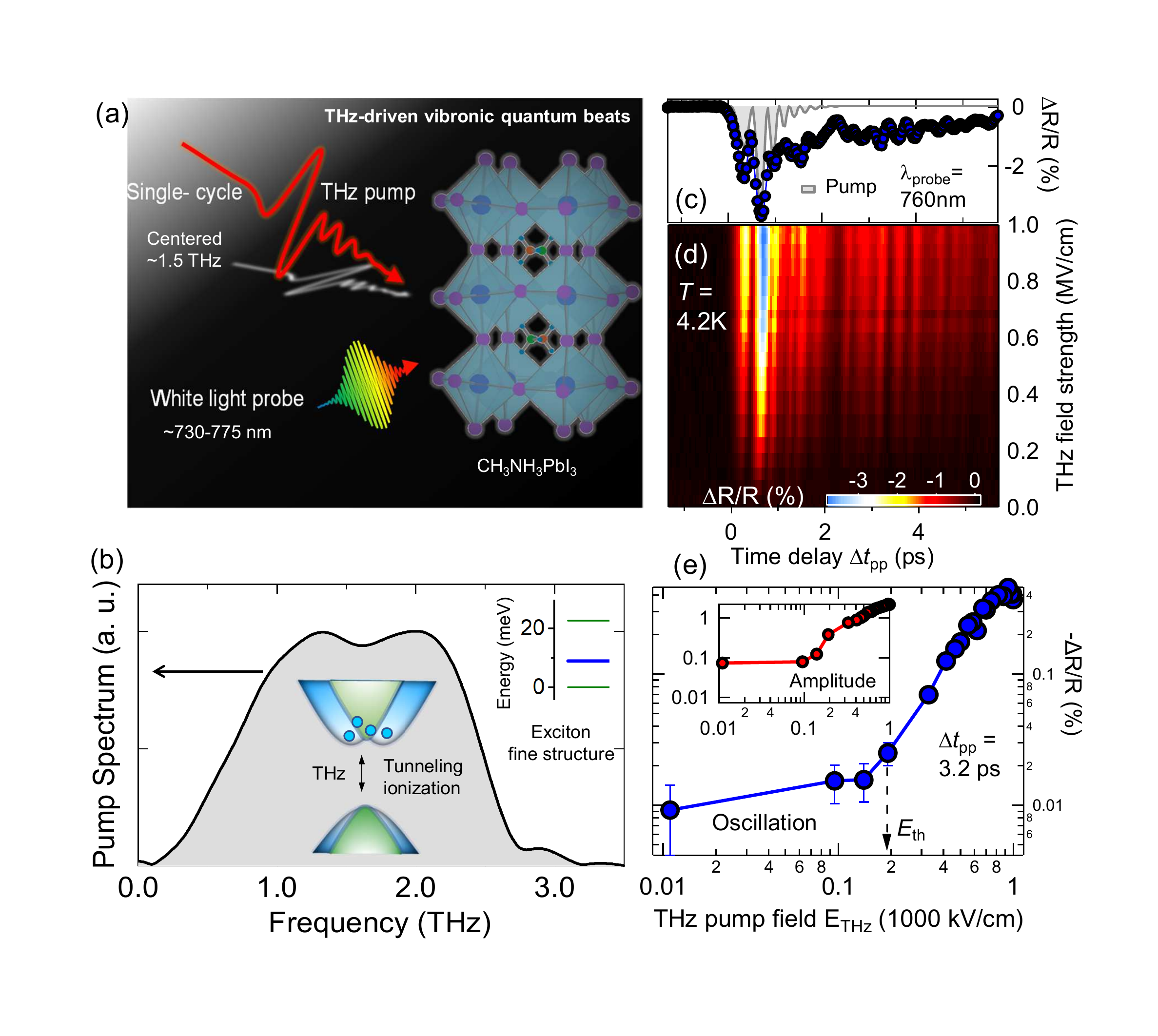}
	\end{center}
	\caption{(a) Schematics for THz pump and white light continuum probe spectroscopy in MAPbI$_3$.  
		(b) THz pump spectrum (gray shade) shown together with an illustration of Rashba-type bands. 
		(c) Representative $\Delta R/R$ dynamics at 760 nm probe induced by E$_{THz}=$938 kV/cm 
	 	at 4.2K and (d) complete 2D false-colour plot. Squared THz pump field is plotted together in (c) (gray shade).
	 	(e) Pump electric field dependence of amplitudes of the oscillatory (blue circles) and exponential decay (inset, red circles) components. See supplementary for a detailed analysis (Fig. S3). } 
	\end{figure}

Coherent quantum beat spectroscopy driven by an intense single-cycle terahertz (THz) pulse, Fig. 1(a),
is extremely relevant for discovery and control of the excitonic FSS. 
First, the deep sub-bandgap pump spectrum centered at $\sim$6~\,meV or 1.5 THz (gray, Fig. 1(b)) can lead to 
coherent excitation of phonons \cite{phonon} with minimum heating.
Such low energy, non-thermal pumping allows the determination of the genuine ground state, with only few nm splitting, 
complementary to, e.g., 
high energy, optical pumping \cite{Miyata, optical,Thouin}.  
Second, in addition to the conventional THz phonon generation mechanism from photonic or ionic coupling \cite{phonon}, polaronic $e$-phonon coupling in perovskites allows the generation of vibronic quantum beats from the intense THz pulse-driven tunneling ionization.
Microscopically, this creates transient exciton population mediated by interband exciton coherence which, in turn, impulsively distorts vibrational potential and generates exciton band oscillations dressed by coherent phonons. 
Third, 
the driven ``ultrafast-ultrasmall" lattice displacement senses the local dynamic dielectric environment and modulates excitons that can be probed by quantum beats. 
The dark states can be periodically brightened by broken symmetry coherent phonons, unlike thermal ones averaging to zero. 
This may determine the FSS by their {\em symmetry-selective} coupling to infrared (IR) or Raman modes that strongly modulate the spin-split bands. 
Such unique coherent phenomenon has never been observed. 

Here we report vibronic quantum beats from the Rashba-type fine structure induced by an intense single-cycle THz field in bulk MAPbI$_3$ crystals. This manifests as {\em symmetry-selective coupling} of dark and bright excitons to predominantly Raman and IR phonon coherence. 
Fig. 1(c) shows 
a pronounced oscillatory behavior with complex beating patterns superimposed on a slow amplitude decay in the pump-induced differential reflectivity $\Delta R/R$ dynamics at 760 nm probe wavelength, i.e., the exciton peak at 4.2\,K, after the sub-ps pump (gray shade) at E$_{THz}=$938 kV/cm.
A complete 2D false-colour plot of $\Delta R/R(\Delta t_{pp}, E_{THz})$ is shown in Fig. 1(d) as a function of time delay $\Delta t_{pp}$ up to 5.7  ps and $E_{THz}$ pump up to $\sim$1000\,kV/cm. 
This allows to extract the amplitudes associated with the oscillatory (Fig. 1(e)), and exponential decay components (inset).   
Intriguingly, the THz field dependence shows a clear nonlinearity with a threshold-like turning on of a new quantum process, i.e., the critical feature for the tunneling ionization with a threshold E$_{th}\sim$200 kV/cm in the beating signals. 
A saturation of excitonic signals is seen at E$_{sat}\sim$1000 kV/cm.  
Note that the vibronic quantum beats here exhibit much more complex patterns than previously observed in perovskites, e.g., up to 100 kV/cm \cite{Kim}. 
One order of magnitude higher THz field strength used \cite{Yang, Xu1, Liang1, npj,lightwave} and long-lived charge carrier dynamics after the pulse, from the transient population decay, clearly identify the roles of THz-driven tunneling ionization.  

\begin{figure}
	\begin{center}
			\includegraphics[width=150mm]{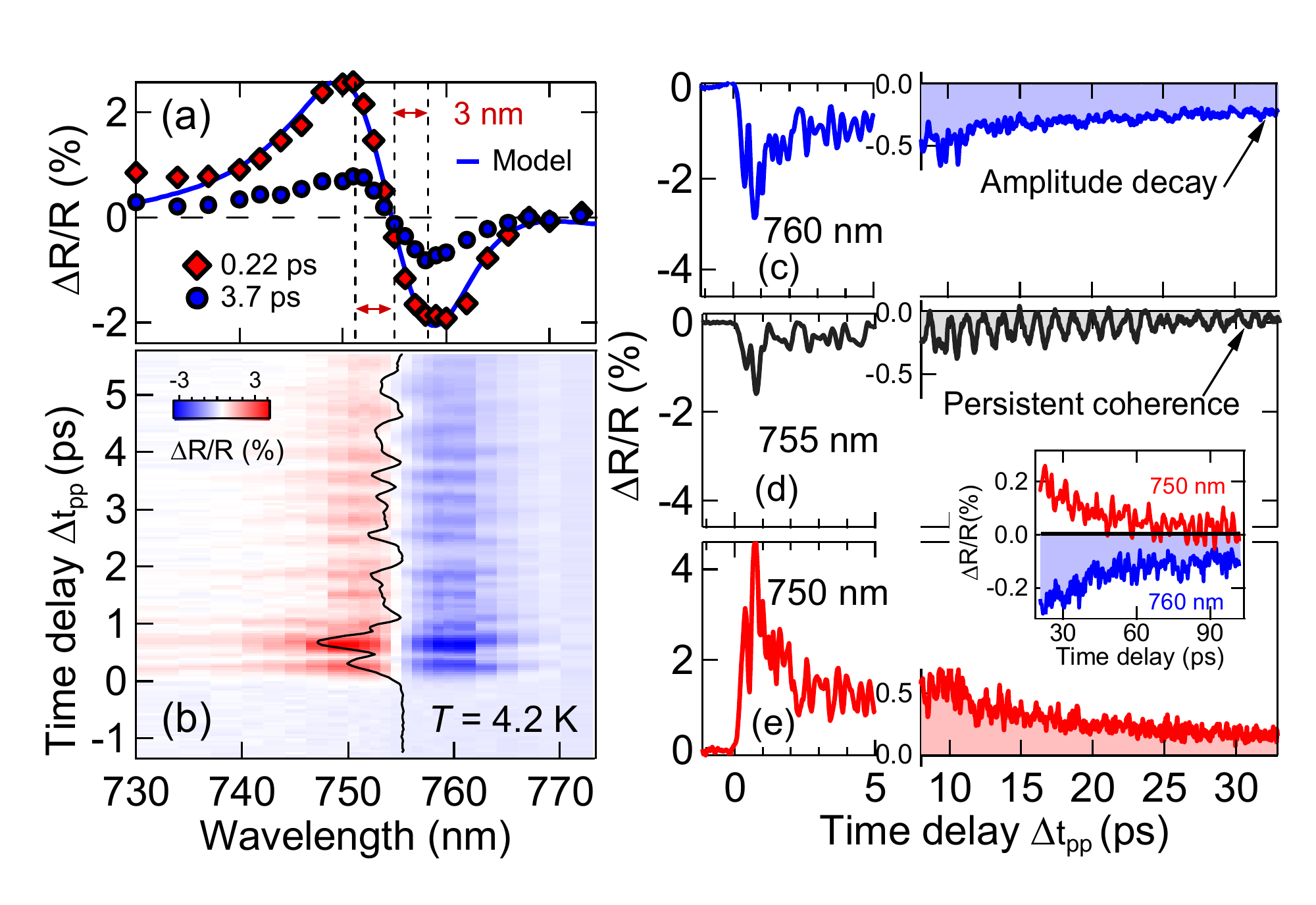}
	\end{center}
\caption{
(a) A time-cut, $\Delta R/R$ spectrum at $\Delta t_{pp}$= 0.22 ps (red) and 3.7 ps (blue) from 2D false-colour plot of THz pump-induced white light spectra in (b).
Three stable spectral positions marked, i.e., positive (negative) peaks and zero crossing. Shown together in (a) is an oscillator model simulation (blue line, see eq. (2) for details).
(c)-(e): The wavelength-cut, three $\Delta R/R$ dynamics are shown at 760 nm (blue), 755 nm (black) and 750 nm (red), respectively. 
Inset of (d) and (e): clearly longer amplitude decay at the 760 nm trace than the 750 nm one up to 100 ps.}
\end{figure} 

The experiment was performed at 4.2 K in a single crystal MAPbI$_3$ sample in its orthorhombic phase. We performed single cycle intense THz pump and white light continuum probe spectroscopy with simultaneous fs temporal and sub-nm spectral resolutions (supplementary). Fig. 2 presents the spectral-temporal behavior of THz pump-induced $\Delta R/R$.  
The broadband probe light ranging from 730 nm to 773 nm is spectrally resolved after the sample and used to probe excitonic states in MAPbI$_3$ after THz excitation $E_{THz}$= 938\,kV/cm. 
The transient spectra exhibit a broad bi-polar, inductive lineshape (Fig. 2(a)) of $\sim$30 nm width, i.e., a positive $\Delta R/R$ change switches to negative at 755 nm. Most intriguingly, the positive ($\sim$751 nm), negative peaks ($\sim$757 nm) and zero crossing ($\sim$755nm) (dashed lines) show negligible spectral shift during the large $\Delta R/R$ 
change, as shown 
at time delay $\Delta t_{pp}$= 0.22 ps (red diamond) and 3.7 ps (blue circle). The much narrower $\sim$3 nm separation marked in Fig. 2(a) is much smaller than the PL linewidth ($\sim$30 nm, Fig S2, supplementary) and spectral shift ($>$100 nm) in the prior excited state studies by optical pumping \cite{Innocenzo,optical}. 
The latter can be explained largely by the exciton peak shift and broadening nonlinearities induced by the presence of photoexcited hot carriers. 
In contrast, the low-frequency, THz pumping data implies a {\em stable}, narrow three-fold structure, only seen by such non-thermal pumping, which is generic and determined mostly by ground state excitonic structure. 

Next we present wavelength-dependent quantum beat spectra and dynamics that support the hidden three-fold FSS structure
inside the inhomogeneously broadened exciton peak. 
First, there exhibit coherent oscillations with similar patterns on the two sides of the $\Delta R/R$ 2D plot plot in Fig. 2(b). Interestingly, very different oscillations are observed in the center within a very narrow linewidth $\sim$3 nm near the zero crossing position of a broad, $\sim$30 nm $\Delta R/R$ response, e.g., a representative time scan trace is shown near the transition point (black line). 
The different beat patterns from the center to the sides become even more clear by comparing three wavelength-cut dynamics, i.e., the middle, 755 nm probe trace near the zero-crossing position (black, Fig. 2(c)) exhibits remarkably long-lasting quantum oscillations shown for the first 5 ps and for the extended time scales up to many tens of ps (split axis), while the 760 nm (blue, Fig. 2(d)) and 750 nm (red, Fig. 2(e)) traces on the two sides near negative and positive peaks (Fig. 2(b)) have much shorter-lived quantum beats mostly within $\sim$5 ps.  This clearly shows that central exciton state is different from the two sides. 

Second, although the two exciton sidebands at 760 nm and 750 nm show a quasi-symmetric temporal beating pattern within $\sim$5 ps, 
their long time $\Delta R/R$ (population) 
signals after dephasing show different relaxation times with clearly longer-lived signal ($>$100 ps) at the lower, 760 nm trace than the 750 nm one (inset, Figs. 2(c) and 2(e)). This clearly shows that two side bands are also different from each other. The longer decay in the 760 nm trace is consistent with the fact that the final transient carrier recombination arises from the lowest bright exciton. 
Therefore, such asymmetry of the amplitude decay on two sides together with the ``sharp" appearance of new oscillations (Fig. 2(d)) exclusively in the middle cannot be accounted by a single exciton oscillator. These distinguishing features are consistent with the periodically brightened, dark states in the center with persistent quantum coherence with a narrow linewidth since they are much less coupled to the dielectric environment. They are different from the bright states on two sides that are expected to exhibit shorter coherence (and much broader linewidth).

Fig. 3 further reveals the distinct wavelength-dependent phonon modes which disclose a {\em symmetry-selective} coupling to the excitonic FSS in MAPbI$_3$. 
The Fourier transformation (FT) spectra of quantum beats are shown for the same three excitonic states at 760 nm (blue), 755 nm (black) and 750 nm (red) in Figs. 2(c)-2(e), respectively.
The low-frequency Raman spectrum (gray shade) from the same conditions is shown together to identify the mode symmetry.   
The FT spectrum at the center band clearly displays a pronounced peak at $\omega_\mathrm{R1}=$0.8 THz that matches very well with the dominant transverse optical Raman mode from octahedral twist of PbI$_6$ cage \cite{Leguy}. Besides the main peak, there exist multiple secondary peaks mostly below 2 THz that correlate with the THz Raman modes $\omega_\mathrm{R2}$--$\omega_\mathrm{R4}$ (red lines) \cite{note}.  
In strong contrast, the $\omega_\mathrm{R}$ mode is strongly suppressed in the FT spectra of the side excitonic bands at 750 nm and 760 nm, which, instead, exhibit multiple strong peaks $\geq$2 THz, quasi-symmetrically present (black dash lines), i.e., $\omega_\mathrm{I1}=$2  THz,  $\omega_\mathrm{I2}=$2.3  THz,  2.66  THz, 3.1 THz, and  $\omega_\mathrm{I3}=$3.4 THz \cite{note}. 
They are absent in the measured THz Raman spectrum (gray shade) but mostly consistent with the reported IR phonon modes \cite{Leguy,Luo}. 
Prior Raman and linear THz spectra of MAPbI$_3$ reveal the phonon symmetry, IR and Raman activities, and electron-phonon coupling \cite{Leguy}. In the range of 1-4 THz relevant for our experiment, the phonon modes can be understood as a coupled motion of the octahedral cages and associated organic cations. 
Most intriguingly, the distinct comparisons for different probe wavelength in Fig. 3 indicate the preferred coupling of the center (side) excitonic bands to the Raman $\omega_\mathrm{R}$ (IR $\omega_\mathrm{I}$) phonons of very different symmetries.
This robust observation again supports the hidden three-fold FSS structure. 
In addition, 
the 750 nm and 760 nm FT spectra don’t scale each other, indicative of two different bright exciton states with slightly asymmetric coupling to the $\omega_\mathrm{I}$ modes \cite{sup}. 
The multi-fold excitonic fine structure from these results show that the middle dark states are brightened periodically by coherent phonons of Raman symmetry (black) and the bright ones are on two sides (blue and red). The narrow bright/dark state splitting of few nm is consistent with the exciton structure inferred in Fig. 2(a).  

\begin{figure}
	\begin{center}
		\includegraphics[width=150mm]{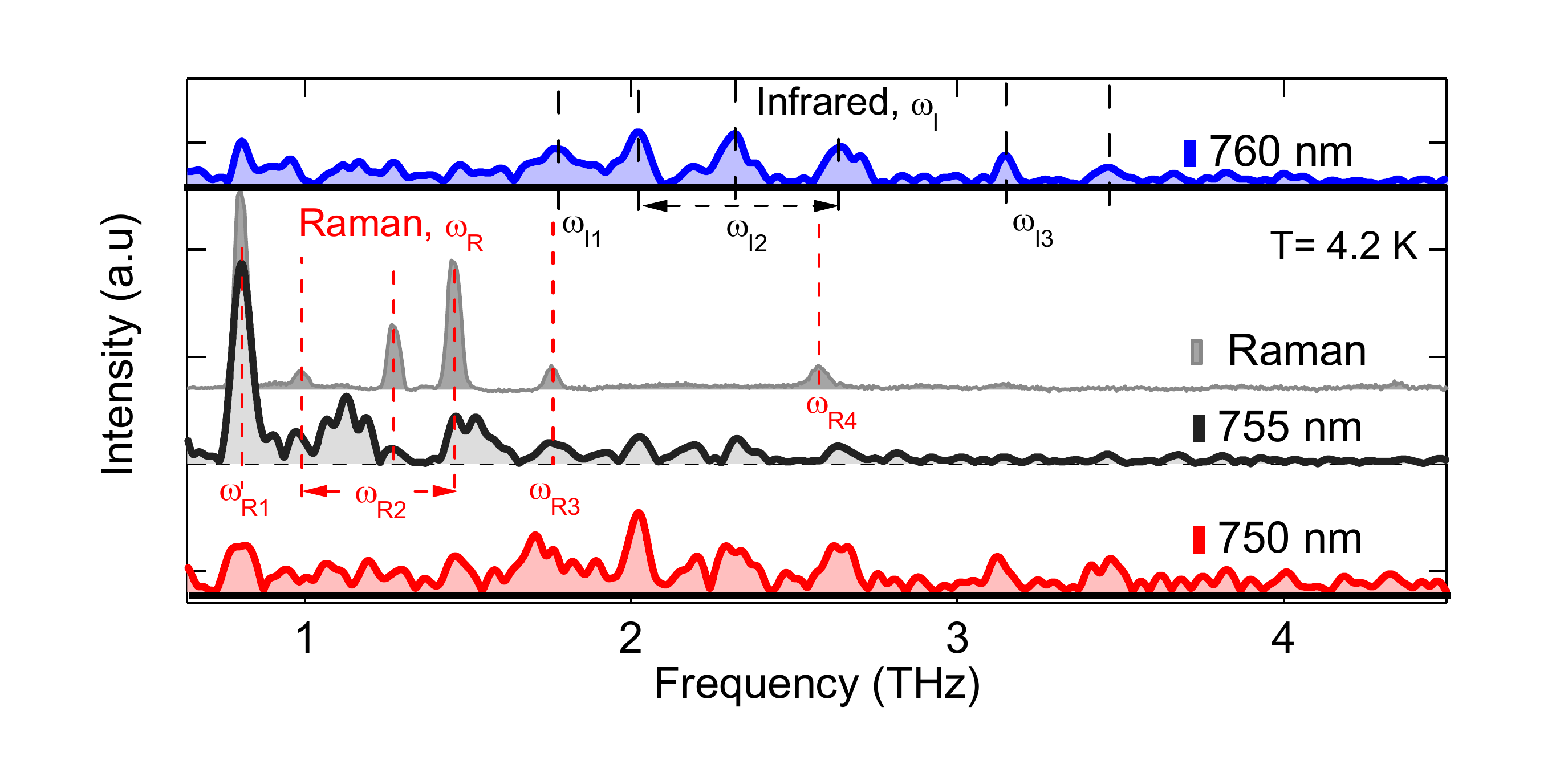}
	\end{center}
\caption{Fourier spectra of three $\Delta R/R$ dynamics in Figs. 2(c)-(e), shown together with the static Raman spectrum (gray shade) 
at 4.2 K. Note the pronounced Raman mode $\omega_\mathrm{R1}\sim$0.8 THz and IR phonon mode $\omega_\mathrm{I1}\sim$2 THz. The assignment and symmetry analysis of other Raman and IR modes are shown in Fig. S6 (supplementary). 
}
\end{figure}  

The time evolution of the {\em symmetry-selective} modes associated with bright and dark states deserves more attention. The quantum beat dynamics at 755 nm (Fig. 2(c)) at 4.2 K for the initial and for the extended 10s of ps time scales (split axis) allow for the determination of the times at which various vibronic modes appear and diminish. At the initial times, all the oscillations at 755 nm also show up in the 760 and 750 nm ones due to the residual IR phonon coupling. These oscillations quickly die out within few ps, indicative a strong vibronic ($e$-phonon) dephasing associated with IR phonons that are coupled with the bright excitons. 
In contrast, at the long times, the $Raman$ band, especially at the $\omega_\mathrm{R1}\sim$0.8 THz, is enhanced by coupling to the middle dark state, which is characterized by a long-lived, tens of ps vibronic coherence, for more than an order of magnitude longer than the bright ones in the 750 and 760 nm traces (Figs. 2(d)-2(e)). This is consistent with the optically-forbidden dark states at the center. 
The exact dephasing mechanism of the virbonic quantum beats is likely from a joint effect of the dark electronic state and Raman phonon symmetry.  
Nevertheless, discovery of the quantum beats clearly reveal the {\em phonon symmetry-selective} coupling to the Rashba fine structure.  

To put the observations 
on a strong footing, we have performed simulations based on an exciton Rashba FSS model, as proposed in Ref. \cite{Isarov}. The exciton FSS due to exchange interaction will be on the order of 0.1 meV for bulk crystal which are neglected in the simualtion \cite{Becker}. The further details are in supplementary. 
The Hamiltonian has the following form in the relative coordinate \cite{sup}
\begin{align}
H=-\frac{\nabla_{r}^{2}}{2\mu}+V(r)+(\alpha_{e}\sigma_{e}-\alpha_{h}\sigma_{h})\cdot\hat{n}\times i\nabla_{r}
\end{align}
where $r=r_{e}-r_{h}$ with $r_{e}(r_{h})$ for the electron (hole) ($e$ and $h$) spatial coordinate. $\nabla$ is the Laplace operator. The reduced mass $\mu$ is defined by $\mu^{-1}=m_e^{-1}+m_h^{-1}$. 
The Rashba term comes from a combined effect of spin-orbit interaction and inversion symmetry breaking (ISB), with ISB field in the $\hat{n}$ direction. 
Although the symmetry breaking source is still debated in perovskites, it can arise at the surfaces and/or bulk local internal structures. 
The optical spectrum of exciton $\Phi$ can be understood by the oscillator strength $f_{\Phi}\propto\frac{\left|P_\Phi\right|^2}{\omega_\Phi}$, with $\omega_{\Phi}$ for exciton oscillator frequency and $P_{\Phi}$ for optical matrix element. The coupling between the exciton and phonon modes is investigated through phonon-modulated model parameters, including $\mu, \alpha_{e}, \alpha_{h}$ and $P_{\Phi}$. They exhibit sinusoidal modulations in time domain following the frozen phonon-induced atomic movement, as shown by DFT calculations (see supplementary section \cite{sup}). Therefore, model simulations reported next use parameters in sinusoidal time modulations with amplitude from these DFT results. 

\begin{figure}
	\begin{center}
			\includegraphics[width=150mm]{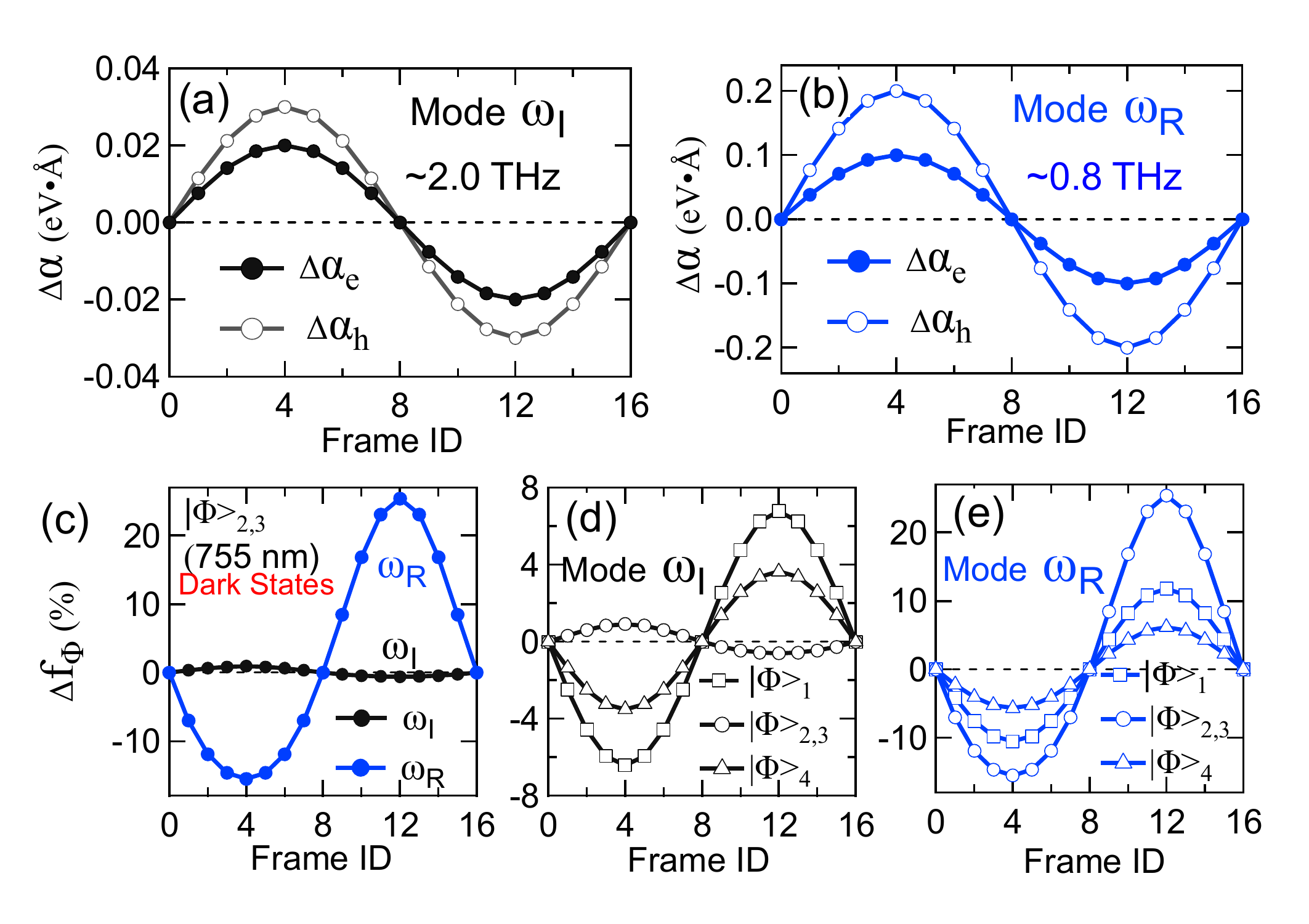}
	\end{center}
\caption{Model calculation for the Rashba states induced by phonon modulation of IR and Raman symmetries. 
	(a) and (b): Changes of the Rashba parameters 
	under $\omega_\mathrm{I1}$ and $\omega_\mathrm{R1}$ vibrations.   
	(c) Oscillator strength changes calculated for the middle, two dark exciton states $ \left|\left.\Phi\right\rangle\right._{2,\ 3}$ for 16 frames during one cycle of the lattice vibration. (d)-(e) Mode-selective coupling to exciton FSS for $\omega_\mathrm{R1}$ and $\omega_\mathrm{I1}$ modes. 
}
\end{figure}  

The lowest energy solution of the above exciton Hamiltonian gives 1{\em s} exciton without Rashba interaction, which has four fold degeneracy due to the electron doublet and hole doublet \cite{Isarov}. With the inclusion of non-zero Rashba term in MAPbI$_3$, the quartet ground state splits to a dark doublet, $\left |  \Phi   \right \rangle_{2,3}$, sandwiched between two bright states, $\left |  \Phi   \right \rangle_{1}$ and $\left |  \Phi   \right \rangle_{4}$ as marked in Fig. 1(b) (inset). Note, however, that the dark doublet is not entirely optically inactive because some amount of atomic orbital character mixing in the $e$ and $h$ states produce relatively small yet finite dipole matrix elements. The ratio is 1:0.1:2.2 for the three-fold FSS states from our calculations \cite{sup}. 

When different phonon modes in the system are excited by THz pump, the bright and dark exciton states respond in different manners. This is predicted by simulating snapshots of Rashba parameters and electronic transitions during one phonon cycle of a specifically-tailored Raman ($\omega_\mathrm{R1}$ mode $\sim$0.8 THz) and IR ($\omega_\mathrm{I1}$ mode $\sim$2.0 THz) symmetries. This shows subcycle brightening of dark excitons by periodically modulating the Rashba parameters and spin-split bands that are fully consistent with our observations. 
In Fig. 4, we plotted the coupling between the four lowest energy exciton states and the two most pronounced phonons identified, Raman $\omega_\mathrm{R1}$ and IR $\omega_\mathrm{I1}$ modes. By comparing the periodical control of the Rashba parameters under these two modes in Figs. 4(a) and 4(b), we clearly notice that the $\omega_\mathrm{R1}$ has much bigger modulation to $\alpha _{e}$ and $\alpha _{h}$. We attribute such distinct enhancement to the facts that the $\omega_\mathrm{R1}$ mode introduces a much larger twist to the PbI$_6$ cage than the $\omega_\mathrm{I1}$ mode and that the Pb-I frame directly determines the electronic band structure of MAPbI$_3$ near the Fermi level. As a result, it is clearly visible, that the $\omega_\mathrm{R1}$ mode changes the bandgap and oscillator strength of the dark states $\left |  \Phi   \right \rangle_{2,3}$ significantly, which gives rise to a pronounced $\omega_\mathrm{R1}$ phonon modulation (blue circles), based on the model calculation in Fig. 4(c), i.e., subcycle coherent brightening of the dark excitons. In contrast, there is nearly no response to the $\omega_\mathrm{I1}$ mode of the same amplitude (black circles). This is in excellent agreement with the experimental observations in Fig. 3. Moreover, we summarize the simulated results for the phonon amplitude dependence of the oscillator strength for the excitonic FSS states when coupled with $\omega_\mathrm{I1}$ (Fig. 4(d)) and $\omega_\mathrm{R1}$ (Fig. 4(e)) modes, respectively. 
For the $\omega_\mathrm{I1}$ coupling in Fig. 4(d), the bright states $\left |  \Phi   \right \rangle_{1,4}$ (rectangle and triangle) exhibit bigger responses than the dark states $\left |  \Phi   \right \rangle_{2,3}$ (circles). The trend is clearly reversed for the $\omega_\mathrm{R}$ coupling in Fig. 4(e). These simulations clearly explain the key mode-selective coupling data in Fig. 3.
 
Finally, we show that the experimentally observed $\Delta R/R$ spectral shape (red diamond) in Fig. 2(a) can be fitted well by two bright oscillators (blue line) centered at $\omega^{}_{01}$=750.9 and $\omega^{}_{02}$=756.7 nm, which correspond to the positive and negative peaks (dash lines). 
Specifically, the dielectric response can be modeled by (supplementary) 
\begin{eqnarray}  \label{equ}
\begin{aligned}
\widetilde{\varepsilon}(\omega)=1+\sum_{j=\text{1,2}}\frac {A_{0j}} {\omega^{2}_{0j}-\omega^{2}+i\omega\tau_{j}^{-1}}. 
\end{aligned}
\end{eqnarray}
Pump-induced transient exciton generation affects $\Delta R/R$ signals by modifying $\widetilde{\varepsilon}(\omega)$ via three main nonlinearities, i.e., bleaching (amplitude A$_{j}$ decrease), broadening (scattering time $\tau_{j}$ decreases) and band gap renormalization (BGR) shifts. The best fit parameters can be found in supplementary. Note that the small BGR shifts $\Delta\omega^{}_{0j}=$0.1 and 0.3 nm are within the spectral resolution, consistent with the stable spectra seen  in Fig. 2(a). 
In addition, the narrow, central dark states will not affect the transient spectra which are determined by the bright states. These results clearly show a Rashba splitting of the bright-dark states $\sim$3nm. This can only be resolved by the unique symmetry-selective, vibronic quantum beats and coherent dynamics shown in Figs. 2 and 3. 

In conclusion, we use THz-driven quantum beat spectroscopy to unambiguously reveal the Rashba-type fine structure in MAPbI$_3$. This inspires ``Rashba engineering"
, i.e., periodic brightening of the dark excitons via modulation of Rashba parameters, for probing and controlling charge transfer and collection. 
Our results also provide compelling implications for merging quantum control \cite{4}, ultrafast phononics and spintronics \cite{1,2,3, 5} to explore multi-function perovskite devices.     

\begin{acknowledgments}
This work was supported by the Ames Laboratory, the US Department of Energy, Office of Science, Basic Energy Sciences, Materials Science and Engineering Division under contract \#DE-AC02-07CH11358 (project supervision, quantum beat spectroscopy and DFT simulations). 
THz Instrument was supported in part by the National Science Foundation Award No. 1611454. 
Sample development at University of Toledo was supported by National Science Foundation DMR 1400432.
Theoretical work at USTC (J.Z. and Y.S.) was supported by National Natural Science Foundation of China under contract number 11620101003, and National Key R$\&$D Program of China 2016YFA0200604 and 2017YFA0204904.

\normalsize{$^\ast$Equal Contribution}
\normalsize{$^\dag$E-mail: jgwang@ameslab.gov}
\end{acknowledgments}



\begin{thebibliography}{99}
\bibitem{Stranks} S. D. Stranks and P. Plochocka, Nat. Mater. \textbf{17}, 381 (2018).
\bibitem{Becker} M. A. Becker, R. Vaxenburg, G. Nedulcu, P. C. Sercel, A. Shabaev, M. J. Mehl, J. G. Michopoulos, S. G. Lambrakos. N. Bernstein, J. L. Lyons, T. St$\ddot{o}$ferle, R. F. Mahrt, M. V. Kovalenko, D. J. Norris, G. Rain$\grave{o}$, and A. L. Efros, Nature \textbf{553}, 189 (2018).
\bibitem{PRB} Z. Liu et al., Phys. Rev. B, in press (2020) Preprint at https://arxiv.org/abs/2002.08283 (2019) 
\bibitem{Zheng} F. Zheng, L. Z. Tan, S. Liu, and A. M. Rappe, Nano Lett. \textbf{15}, 7794 (2015).
\bibitem{Giovanni} D. Giovanni, H. Ma, J. Chua, M. Gr$\ddot{a}$tzel, R. Ramesh, S. Mhaisalkar, N. Mathews, and T. C. Sum, Nano.Lett. \textbf{15}, 1553 (2015).
\bibitem{spin} P. Odenthal, W. Talmadge, N. Gundlach, R. Wang, C. Zhang, D. Sun, Z. -G. Yu, Z. V. Vardeny and Y. S. Li, Nat. Phys. \textbf{13}, 894 (2017).
\bibitem{K} K. Frohna \textit{et al.} \emph{Nat. Commun.} \textbf{9}, 1829 (2019).
\bibitem{Isarov} M. Isarov, L. Z. Tan, M. I. Bodnarchuk, M. V. Kovalenko, A. M. Rappe, and E. Lifshitz, Nano Lett. \textbf{17}, 5020 (2017).
\bibitem{Liu} Z. Liu, K. C. Bhamu, L. Liang, S. Shah, J. -M. Park, D. Cheng, M. Long, R. Biswas, F. Fungara, R. Shinar, J. Shinar, J. Vela, and J. Wang, MRS Commun. \textbf{8}, 961 (2018).
\bibitem{Beecher} A. N. Beecher, O. E. Semonin, J. M. Skelton, J. M. Frost, M. W. Terban, H. Zhai, A. Alatas, J. S. Owen, A. Walsh, and S. J. L. Billinge, ACS Energy Lett. \textbf{1}, 880 (2016).
\bibitem{Innocenzo} V. D$^,$Innocenzo, G. Grancini, M. J. P. Alcocer, A. R. S. Kandada, S. D. Stranks, M. M. Lee, G. Lanzani, H. J. Snaith, and A. Petrozza,  Nat. Commun. \textbf{5}, 3586 (2014).
\bibitem{Milot} R. L. Milot, G. E. Eperon, H. J. Snaith, M. B. Johnston, and L. M. Hertz, Adv. Funct. Mater. \textbf{25}, 6218 (2015).
\bibitem{Tom} T. J. Savenije, C. S. Ponseca, L. Kunneman, M. Abdellah, K. Zheng, Y. Tian, Q. Zhu, S. E. Canton, I. G. Scheblykin, T. Pullerits, A. Yartsev, and V. Sundstr$\ddot{o}$m, J. Phys. Chem. Lett. \textbf{5}, 2189 (2014).
\bibitem{Fu} M. Fu, P. Tamarat, H. Huang, J. Even, A. L. Rogach, and B. Lounis, Nano Lett. \textbf{17}, 2895 (2017).
\bibitem{wilhelm} D. Niesner, M. Wilhelm, I. Levgen, A. Osvet, S. Shrestha, M. Batentschuk, C. Brabec, and T. Fauster, Phys. Rev. Lett. \textbf{117}, 126401 (2016).
\bibitem{JPC} D Cheng, et al., The J. Chem. Phys. \textbf{151}, 244706 (2019).
\bibitem{phonon} M. Kozina, M. Fechner, P. Marsik, T. van Driel, J. M. Glownia, C. Bernhard, M. Radovic, D. Zhu, S. Bonetti, U. Staub, and M. C. Hoffmann, Nat. Phys. \textbf{15}, 387 (2019). 
\bibitem{Miyata} K. Miyata, D. Meggiolaro, M. T. Trinh, P. P. Joshi, E. Mosconi, S. C. Jones, F. De Angelis, and X. -Y. Zhu, Sci. Adv. \textbf{3}, e1701217 (2017).
\bibitem{optical} T. Ghosh, S. Aharon, L. Etgar, and S. Ruhman, J. Am. Chem. Soc. \textbf{139}, 18262 (2017).
\bibitem{Thouin} F. Thouin, D. A. Valverde-Ch$\acute{a}$vez, C. Quarti, D. Cortecchia, I. Bargigia, D. Beljonne, A. Petrozza, C. Silva, and A. R. Srimath Kandada, Nat. Mater. \textbf{18}, 349 (2019).
\bibitem{Kim} H. Kim, J. Hunger, E. C$\acute{a}$novas, M. Karakus, Z. Mics, M. Grechko, D. Turchinovich, S. H. Parekh, and M. Bonn, Nat. Commun. \textbf{8}, 687 (2017).
\bibitem{Yang} X. Yang, C. Vaswani, C. Sundahl, M. Mootz, P. Gagel, L. Luo, J. H. Kang, P. P. Orth, I. E. Perakis, C. B. Eom, and J. Wang, Nat. Mater. \textbf{17}, 586 (2018).
\bibitem{Xu1} Yang, X. \textit{et al.} \emph{Nat. Photonics.},  \textbf{13}, 707 (2019).
\bibitem{Liang1} Liang, L. \textit{et al.} \emph{Nat. Commun.} \textbf{10}, 607 (2019).
\bibitem{npj} X. Yang et al., \textit{npj Quantum Mater.}, \textbf{5}, 13 (2020). https://doi.org/10.1038/s41535-020-0215-7
\bibitem{lightwave} Vaswani, C. et al., Preprint at https://arxiv.org/abs/1912.01676 (2019).
\bibitem{Leguy} A. M. A. Leguy, A. R. Go$\tilde{n}$i, J. M. Frost, J. Skelton, F. Brivio, X. Rodr{\'i}guez-Mart{\'i}nez, O. J. Weber, A. Pallipurath, M. I. Alonso, M. Campoy-Quiles, M. T. Weller, J. Nelson, A. Walsh, and P. R. F. Barnes, Phys. Chem. Chem. Phys. \textbf{18}, 27051 (2016).
\bibitem{note} For the detailed assignment of Raman and IR phonons and their motions, see Fig. S6 in supplementary.  
\bibitem{Luo} L. Luo, L. Men, Z. Liu, Y. Mudryk, X. Zhao, Y. Yao, J. M. Park, R. Shinar, J. Shinar, K. M. Ho, I. E. Perakis, J. Vela, and J. Wang, Nat. Commun. \textbf{8}, 15565 (2017).
\bibitem{sup} See Supplemental Material [url] for further technical details of samples used, experimental setup and model, which includes
Refs. [36-40].
\bibitem{4} Vaswani, C. et al., Phys. Rev. X, in press (2020); Preprint at https://arxiv.org/abs/1912.02121 (2019).
\bibitem{1} T. Li, et al., Nature {\bf 496}, 69 (2013) 
\bibitem{2} Yang, X. et al., \textit{Phys. Rev. Lett.} \textbf{121}, 267001 (2018)
\bibitem{3} Patz, A. et al., \textit{Phys. Rev. B.} \textbf{91}, 155108 (2015)
\bibitem{5} A Patz, T Li, L Luo, X Yang, S Bud'ko, PC Canfield, IE Perakis, J Wang, \textit{Phys. Rev. B} \textbf{95}, 165122 (2017) 
\bibitem{ref36}M. I. Saidaminov, A. L. Abdelhady, B. Murali, E. Alarousu, V. M. Burlakov, W. Peng, I. Dursun, L. Wang, Y. He, G. Maculan, A. Goriely, T. Wu, O. F. Mohammed, and O. M. Bakr, Nat. Commun. \textbf{6}, 7586 (2015).
\bibitem{ref37}B. Adolph, J. Furthmüller, and F. Bechstedt, Phys. Rev. B \textbf{63}, 125108 (2001).
\bibitem{ref38} G. Kresse and D. Joubert, Phys. Rev. B \textbf{59}, 1758 (1999).
\bibitem{ref39} J. P. Perdew, K. Burke, and M. Ernzerhof, Phys. Rev. Lett. \textbf{77}, 3865 (1996).
\bibitem{ref40}S. Grimme, J. Comput. Chem. \textbf{27}, 1787 (2006).
  


\end{thebibliography}
\end{document}